# When Energy Trading meets Blockchain in Electrical Power System: The State of the Art

**Naiyu Wang[1], Xiao Zhou[1], Xin Lu[1], Zhitao Guan[1, *], Longfei Wu[2], Xiaojiang Du[3] and Mohsen Guizani[4]**

[1] School of Control and Computer Engineering, North China Electric Power University, Beijing, China; shininess_y@163.com
[2] Department of Mathematics and Computer Science, Fayetteville State University, Fayetteville, NC, USA; lwu@uncfsu.edu
[3] Department of Computer and Information Science, Temple University, Philadelphia PA, USA; dxj@ieee.org
[4] Electrical and Computer Engineering Department, University of Idaho, Moscow ID, USA; mguizani@gmail.com
* Correspondence: guan@ncepu.edu.cn

**Abstract:** With the rapid growth of renewable energy resources, the energy trading began to shift from centralized to distributed manner. Blockchain, as a distributed public ledger technology, has been widely adopted to design new energy trading schemes. However, there are many challenging issues for blockchain-based energy trading, i.e., low efficiency, high transaction cost, security & privacy issues. To tackle with the above challenges, many solutions have been proposed. In this survey, the blockchain-based energy trading in electrical power system is thoroughly investigated. Firstly, the challenges in blockchain-based energy trading are identified. Then, the existing energy trading schemes are studied and classified into three categories based on their main focus: energy transaction, consensus mechanism, and system optimization. And each category is presented in detail. Although existing schemes can meet the specific energy trading requirements, there are still many unsolved problems. Finally, the discussion and future directions are given.

## 1. Introduction

Energy crisis and environmental pollution have become two critical concerns in the last decade [1]. Smart grid with renewable energy resources is a promising approach to alleviate these problems, and its implementation can be facilitated by microgrids. In smart grid, new models, technologies, and flexible solutions are to be developed for intelligent energy management, to facilitate the optimal energy and power flow strategies. Specifically, energy trading, as an important component in energy management, is undergoing an evolution from centralized to distributed manner. In traditional energy market, the energy users can only buy electricity from the power utilities. In Smart grid, the energy users play a role in both power consumer and supplier. The excessive renewable energy generations can be traded with the utility and other users in deficit of power supplies for mutual benefits.

The traditional energy market model has many problems, such as over-centralization, low efficiency, high energy cost, security and privacy issues and so on, and its development has met a bottleneck [2-5]. Additionally, in the transition from traditional power grid to smart power grid, the efficiency, demand, cost and emission issues also need to be fully considered [6].

Meanwhile, as an emerging technology, blockchain has been widely adopted in many fields for its well-known advantages [7,8]. Blockchain was indeed flawed when it first emerged.

Nevertheless, studies have found these problems and made corresponding improvements, such as the measurement research and exploration of the re-parameterization of blockchain to improve its scalability [9], the bitcoin-NG [10] which can tolerate Byzantine fault tolerance and achieve very good scalability, and the Protocol ''Ouroboros Praos'' which can defend against complete adaptive corruption [11].

The application fields of the blockchain technology include finance, Internet of things, public social services, reputation system, security and privacy, and so on [12]. For example, in [13], blockchain is used to enhance privacy through a decentralized personal information management system. It can ensure that users have control over data, proving that blockchain may serve as an important tool for trusted computing. The Hawk model proposed in [14] also addresses the issue of transaction privacy, by using an intelligent contract system that enables the compiler to automatically generate an efficient encryption protocol. As for the Internet of Things, blockchain can enable the sharing of services and resources, and automate several existing time-consuming workflows in the form of cryptographic authentication [15]. Beekeeper, a secure blockchain-based threshold IoT service system was proposed in [16]. Blockchain can also be employed for secure decentralized energy management [17]. Another study [18] has used blockchain to design privacy-protected search scheme for malicious servers. Moreover, it can also be applied to data sharing schemes to ensure the fairness of incentives [19].

The integration of blockchain technology into energy trading is a novel and promising area of research, and many studies have made efforts in this regard [20]. Before formally attempting to combine blockchain with the energy trading model, some researchers first studied the application of distributed systems to the energy market. Proof by facts, P2P network model enables the energy market to operate in a consumer-centered manner and support the participation of the so-called "prosumers", which has greatly improved the flexibility of the traditional mode of energy market [21-24]. Additionally, the distributed system can also provide high-precision demand response signals [25], reduce cost and improve speed [26]. Moreover, it solves the problem of system scalability and mobility, and creates a competitive market that benefits small-scale prosumers [27]. Other studies mainly focus on the coordination between the macro power grid and micro grids, between micro grids and renewable resources, and within individual power grids, to formulate strategies that consider user utility, improve resource utilization, and ultimately achieve cost minimization [28-30].

Many studies have shown the advantages of introducing blockchain technology into the energy trading systems [31,32]. For example, high computing capacity can be realized with blockchain at a low cost, and the consensus mechanism can ensure the optimal solution. In addition, it can prevent fake transactions and establish an open and transparent credit system [33]. All nodes can work independently towards system equilibrium without central supervision [25]. The application of cryptography in blockchain can solve a number of system security problems [27,34] and significantly reduce costs [35]. More generally, blockchain enhances the role of consumers in the system and their choices [35]. In [36], the authors have evaluated the local electricity market based on blockchain, and the model shows that the system has already been in the early maturity stage.

Based on the survey of existing research work, this paper firstly expounds the problems faced by energy transaction before and after the adoption of blockchain. Then, we will analyze the relevant researches on combining energy transaction with blockchain. Next, existing research are summarized and classified based on different aspects in the blockchain model. Finally, we summarize and highlight the problems and challenges that still need to be addressed in this field.

The rest of this paper is organized as follows. Section II gives the preliminaries. In section III, the challenges of blockchain-based energy trading are stated. In section IV, an overview of blockchain-based energy trading is given. In section V, the methodologies for energy transactions are investigated. The consensus mechanisms are studied in Section VI. In Section VII, the methodologies for system optimization are presented in detail. In Section VIII, the discussion and future directions are stated. In Section X, the paper is concluded.

## 2. Preliminaries

### 2.1. Blockchain

As the underlying technology of bitcoin [7], blockchain adopts a decentralized system framework with the proof-of-work consensus institution. It can avoid the double-spending problem. The transactions are linked together by the secret keys and timestamps to form a chain.

When new transactions are generated and broadcasted, each node collects the new transactions into a block. Then, they work on finding a difficult proof-of-work for its block. When someone finds an answer, it broadcasts the block to all nodes. All other nodes check if it is valid and has not been spent.

In addition, an appropriate incentive mechanism is proposed to encourage more miners to join the system and run the mining computations. After the great success of the application in bitcoin, blockchain has been explored in other possible applications with the addition of smart contracts.

### 2.2. Zero-knowledge proof

A zero-knowledge proof is a convincing proof to someone that "I do own something without revealing the solution or the exact content of what I own" [37,38].
- Interactive zero-knowledge proof. In the presence of both the prover and the verifier, the prover claims to possess the knowledge and the verifier keeps challenging the prover until the responses from the prover can convince the verifier that the prover does possess the claimed knowledge. But if the two parties collude, they could spoof others into believing that they possess the knowledge without actually knowing it.
- Non-interactive proof of zero knowledge. The interactions between prover and verifier is not necessary. Instead, an additional machine program that no one knows about will be used. The proof calculated by the machine automatically prevents either party from cheating.

### 2.3. Ring signature

Ring signature scheme makes use of algorithms to construct signatures with unconditional anonymity. It is mainly composed of the following algorithms [39-41]:
- Generate. A probabilistic polynomial time (PPT) algorithm, with security parameter k as input, uses different public key systems to generate public and private keys for each user and output the result .
- Sign. When signing the message m, the public key of n ring members and the private key of one of them will be used to generate a signature R with ring parameters.
- Verify. A deterministic algorithm that takes a message and a signature as inputs. If the message matches the signature, it returns 'true'; otherwise, it returns 'false'.

### 2.4. Cloud Computing & Fog Computing

Cloud Computing is a computing paradigm in which the service providers can remotely distribute computing and storage resources to users through the networks in an on-demand manner. The resources are available to users without their direct active management [42].

Fog Computing takes advantages of weaker and decentralized edge devices in place of the powerful servers in cloud Computing. Fog computing has a closer proximity to users and larger geographical distribution. [43].

### 2.5. Onion routing & garlic routing

Onion routing is a technique for anonymous communication. In an onion network, messages are encapsulated in multiple layers of encryption like an onion. The encrypted message will pass through a series of onion routers, each of which decrypts one single layer, uncovering the next node [37]. Only the original sender knows where the destination is, and only the original sender and the destination node know the plaintext message. All intermediate nodes know only the immediately preceding and following nodes, hence achieving the anonymity of the original sender and the message.

Garlic routing is a variant of onion routing. Different to onion routing, it groups messages and disperses them through multiple tunnels, which further increases the difficulty of attack.

## 3. Challenges of Blockchain-based Energy Trading

Disadvantages of traditional energy trading model are as followed:
i.  Centralized third-party control: as an important part of the central network, errors of the centralized third-party will interfere with the authentication and payment activities, causing severe damage to the operation and security of the whole system [44,45]. In addition, the centralized system management leads to high operating costs, low transparency, and potential risk of transaction data modification [46]
ii. Privacy protection and security issues. Third parties may disclose an user's energy generation pattern and predict an user's daily activities and behavior rules based on historical information.
iii. It does not reflect the cause-and-effect relationship between the expense of each consumer for the use of the grid and the total energy cost associated with the grid [47].
iv. The lack of competition between renewable energy pricing and traditional market pricing will discourage the investment in renewable energy [47].
v. The grid lacks the resilience to cyber-attacks [48].
vi. Energy is generated from centralized and large energy plants, which reduces the flexibility to adapt to an increasing demand [49].
vii. Energy tends to travel long distances before reaching the consumers, and only one energy flow is supported, making it vulnerable to disruptions [49, 50].

However, blockchain is also vulnerable to attacks. For example, in [55], five attacking methods were identified in the Internet-of-Things system based on blockchain. We summarize the following challenges for the systems combined with blockchain:
i.  Low efficiency [52] and high transaction costs [53,54]. Due to the shortcomings of blockchain itself, the transaction speed cannot meet the system requirements. At the same time, the transaction costs of blockchain can raise the overall cost of the system.
ii. Privacy protection and security issues [52,54,55]. How to avoid big data statistical prediction and behavior model analysis when the transaction is concluded and the rights and interests of both parties to the transaction can be guaranteed may be a severe challenge. There are also potential risks such as private key leak [8].
iii. Real-time communication [53]. The challenge of the real-time communication of a large amount of sensor data.
iv. Price model. How to price a deal properly to help buyers and sellers reach an agreement on quickly?
v. Lack of motivation [8,52]. How to encourage manufacturers (manufacturers of what?) to join the system when there are many privacy concerns.
vi. Cost minimization, multi-scene welfare maximization. In the case of multiple variables, how to ensure the minimization of the overall energy cost of the system, as well as the assurance of trust, rationality of independent individuals, and computational efficiency [56].
vii. Lack of a regulatory framework with flexibility [53,57,31,36].
viii. There is also the issue of environmental energy consumption, because the system requires a large carbon footprint [53].

ix. Because of the tight network structure, there may be a cascade of physical faults [8]. Also, there are many physical hardware limitations to consider [58].

## 4. Overview of Blockchain-based Energy Trading

Currently, the application of blockchain in the field of energy trading has achieved significant success. One of the most common forms is based on P2P transactions, using the blockchain as the system framework. P2P networks can be roughly divided into two organizational modes, structured and unstructured. Each structured peer is built from a distributed hash table, and message routing is efficient but requires maintenance. The unstructured peer location is random, without the need for a centralized node search, but the query may not have a result. Jogunola [50] et. al evaluated both the structured and unstructured P2P protocols to help planners better organize their networks. The conclusion is that both of them are highly scalable while the structured P2P protocols are more reliable by comparison.

As shown in figure 1, this paper takes the overall structure model of blockchain as the context. Start by reviewing each step of the trading process, then consider improvements based on the consensus mechanism, and finally consider macro-optimization of the entire system.

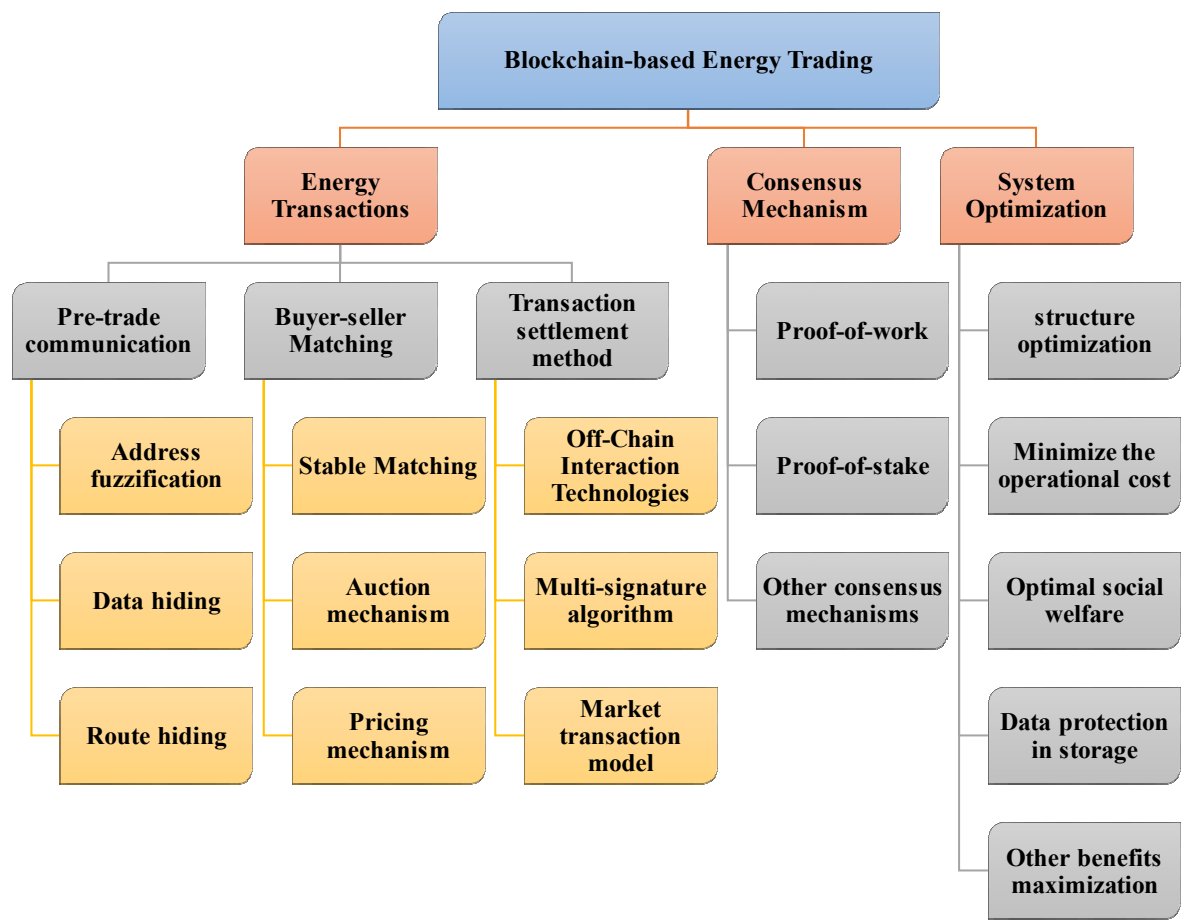

Fig.1 Classification of Blockchain-based Energy Trading

## 5. Category I: Energy Transaction

*5.1. Pre-trade communication*

Users often need to publish their own supply and demand before identifying the trading partners. In this process we need to ensure security and protect users' privacy.

5.1.1. Address fuzzification

First of all, there is a simple way to hide the actual address of the user, anonymous addresses, in which a new pseudonym address is generated for each transaction. Most digital currencies using blockchain technology employ anonymous addresses to provide anonymous services [44]. System forces users to randomly generate new message addresses for each new transaction to remain anonymous [59,60]. However, the connection between the pseudonym and the user can be found through the matching between the electricity consumption data and the user behaviors. Therefore, in [61], each user is allowed to create multiple pseudonyms and submit their power consumption data under different pseudonyms to break such link. Furthermore, bloom filters are used to determine the validity of pseudonyms and zero knowledge is used to verify the existence of pseudonyms.

At the same time, zero knowledge proof also has other applications. For example, in [38], Non-interactive zero-knowledge proof is also used to realize address fuzzification by hiding a coin to the coin list, which can successfully sever the link between the user and the coin. Moreover, the ring signature algorithm mixing through Zero-Knowledge Proofs in [37] is used to generate many new wallets addresses to realize the intractability and unlinkability of transactions.

5.1.2. Data hiding

The basic idea is to hide the data that needs to be hidden into a large data set.

In [62], existing household energy storage units have been used to blur household consumption and reduce the cost of load hiding. The opportunistic joint use of electric vehicles, heating, ventilation and air-conditioning systems can reduce or eliminate the reliance on locally dedicated rechargeable batteries. The mutual information between the original household load and the mask load has been used as the privacy protection measures. For the mask load time series and the original load time series, the combination of existing thermal appliances and energy storage units has been used to hide the household load. In [63], a data confusion scheme is proposed. In the bidding mechanism of power trade, only the information regarding the overall customer transactions is released to suppliers, so that they are able to calculate monthly bills but have no access to fine-grained data. Producers need to transfer production assets to an intermediate mixing service as proposed in [59]. Since the service simultaneously moves assets from multiple producers to multiple anonymous addresses, and the anonymous addresses are randomly generated by the prosumers, the blended assets cannot be traced back to the original prosumer. Similarly, the group signature algorithm adopted in [37] allows members of a group to send transactions in such a way that the receiver can only know that it comes from a group member but nothing more, without using a central authority.

In [64], a virtual power plant is used as the intermediator for the negotiation between prosumers and the aggregator, without sharing any private information. Additionally, in order to prevent the information leakage caused the side-channel attacks during data exchange, a bilinear ECDSA signature mechanism is proposed in [65] against adaptively chosen messages attacks under the continual leakage setting.

5.1.3. Route hiding

The garlic routing scheme proposed in [37] takes intelligent electricity meters, manufacturers and distribution system operators as the server nodes for garlic routing, and applies netDB to realize the routing allocation, thus preventing attackers from obtaining some private information of both parties in a transaction based on the routing messages.

In [54], a routing method using the public keys as identifiers is proposed. That is, nodes with high resource availability are responsible for forming the backbone network. Backbone nodes route groups based on the destination's public key. Normal nodes can be associated with multiple backbone nodes for protection of anonymity with multiple public keys.

*5.2. Matching*

5.2.1. Stable Matching

In order to protect the privacy of fixed residential charging piles, mobile electric cars and other charging equipment, a homomorphic encryption-based position hiding method is proposed in [66] so that the exact locations of the supplier and demanders can be known only if the stable matching is successful. When matched, the mobile suppliers need to send their updated location to the demander. This method generates the preference list in the ascending order of distances, which is calculated from the encrypted location information of the suppliers and demanders. Finally, based on the preference lists, the distributed stable matching algorithm obtains a matching such that every demander and supplier is satisfied.

In [67], the symmetrical allocation problem based on native auction algorithm is used to match the buyers and sellers in the energy market. The naive auction algorithm is carried out in rounds, and only one buyer bids for the desired object in each round.

In [68], both the buyer and seller provide a vector and some matching criteria under the premise of ensuring safety. The objective is to find an optimal set of solutions that satisfy these criteria. With further consideration of time constraints, physical constraints, and more flexible pricing updates, the system can provide the most appropriate matching strategy. – For this paragraph, the main contributions that the authors of [68] think that their paper have made are not mentioned. By reading the technical details in this paragraph, readers will not have a high-level idea of what [68] is working on, which can be found from the abstract, main contributions, and conclusion sections.

5.2.2. Auction mechanism

Initially, the auction mechanisms were used only in a distributed model. For example, in [69], the auction mechanism is skillfully compatible with incentives, and the distributed mechanism is adopted to transfer the auction calculation to users, so as to ensure the honesty of users and reduce the burden on hubs. Moreover, a battery state model for electric vehicles is proposed in [70], which is responsible for clarifying the factors that buyers and sellers need to consider when determining the transaction price and volume during the auction process. The auctioneer determines the optimal allocation by controlling the bidding increment.

When blockchain technology has not been adopted, Majumder [71] et.al proposed an efficient double auction mechanism to maximize the participants' utilities and the social welfare in a distributed system structure, which is completed by the controller in an iterative way, avoiding the need of private information. It is also widely used in distributed grid transactions [58,72,73]. After the blockchain is widely adopted, this auction mechanism's main claim is to match the parties at auction, not the goods to be traded [74,75]. The charging and discharging parties send bidding vectors to the auctioneer respectively, and auctioneer carries out multiple iterative auctions on the price matrix of both parties. The buyer and seller solve their optimal price and the best matching object from multiple iterations. The same mechanism is used in the system and an adaptive aggressive strategy is adopted to adjust the quotation according to the market changes. To prevent unexpected situations, the interests of users can also be guaranteed through multi-signed digital certificates [46]. Furthermore, there are other auction mechanisms that build on top of this. The Double Auction in [76] adopts the average mechanism, in which the reserve price is calculated on an average basis and the seller with a lower reserve price will be allocated to the buyer with a higher bid in a greedy way. In [77], it takes the continuous double auction as the second stage of the auction mechanism. The first stage is the call auction stage, in which buyers and sellers conduct one-time matching transactions and provide guiding prices of various types of energy for each period.

On account of the contradiction and interactivity of energy transaction, the game theory method will be an effective tool to improve the system. The application of game theory in blockchain-based energy trading has been comprehensively studied in [78]. In [79], the non-cooperative game theory -- Nash equilibrium is used to adjust the matching process to achieve the optimal matching. Each player assumes to know the equilibrium strategy of the other player, and users in the system cannot increase their revenue by only modifying their own orders.

There are many other auction mechanisms, such as those algorithms that outsource the mining operations to the Edge Computing Service Provider (ECSP) and maximize the social welfare while guaranteeing the truthfulness [80,81]. In [80], a multi-layer neural network architecture is constructed, and the optimal solution is obtained by deep learning. In the scenario of EV charging [82], a dynamic optimal contract assignment and energy allocation algorithm is devised. Based on contract theory, the optimal contracts are designed to satisfy consumers' distinct energy consumption preferences and maximize the operator's utility. A parallel multidimensional willingness bidding strategy is proposed in [83], to be specific, it is to update the bidding price and give a list of up to three willing trading objects by using historical transaction records, bidding antagonism and other factors between buyers and sellers in the discussion of bidding. In [84], a local auction market was suggested, in which a generic, low-cost solution is implemented using blockchain technology, Specifically, blockchain-based aggregators can perform market clearing, battery energy storage system (BESS) execution and other obligations to ensure that they are applicable to BESS of any size.

5.2.3. Pricing mechanism

The pricing scheme in energy system is very important, it is also related to the system performance. In [85], real-time pricing and direct transactions can be made through the framework of coalitional game theory, and the appropriate electricity price can be easily calculated. A bidding protocol based on secure Multiparty Computation (MPC) is proposed in [63], it allows mutually distrustful parties to compute without disclosing fine-grained electricity consumption data. The trade price and bids selection are performed in a distributed, oblivious manner. The billing protocol is based on a simple privacy-friendly aggregation technique that can prevent the detailed consumption data from leaking to suppliers when they calculate customers' monthly bills

In [86], a novel P2P pricing mechanism is proposed, which can significantly increase the economic benefits of P2P participants and can be used in any P2P energy sharing model. Specifically, a compensating price is introduced to compensate the prosumers to ensure them be better off when they participate in P2P energy sharing. In [75], the zero intelligence plus (ZIP) traders are acting as the agents. Since ZIP traders are subject to budget constraints, they automatically adjust their margins based on the matching of previous orders. A pricing mechanism using the mid-market rate is proposed to ensure the stability of the coalition and the interests of prosumers in [87]. In other words, this is a mechanism to decide whether the electricity needs to be purchased from or sold to the grid according to the relationship between the total power generation and the demand, so as to coordinate the price. In addition to the basic transaction pricing, there are many other aspects to be considered in a pricing mechanism, such as the maximization of system performance.

For example, there are some pricing algorithms based on game theory. Through comparison and evaluation of multiple pricing mechanisms, it is concluded that the pricing mechanisms based on game theory can achieve the best local energy scheduling [88]. In [89], an optimal loan pricing strategy in a credit-based payment scheme is proposed, which makes it possible for buyers who don't have enough coins to buy energy. It also computes the best count of loan for each borrower and the penalty rate through a noncooperative Stackelberg game. Stackelberg game theory is also used in [90,91], in which the service demand pricing of miners is considered from the perspectives of unified pricing and discriminatory pricing respectively, both of which are calculated to determine the best price for profit maximization. 5.3. Settlement

The next step is the process of settling accounts between the two parties of a transaction.

5.3.1 Off-Chain Interaction Technologies

The idea of off-chain payments is similar to creating an escrow account off the chain (More content needed to explain "off the chain": e.g., make multiple transactions without writing into Blockchain) between the two parties of a transaction [92]. Off-Chain Technologies can also help to increase the speed of transaction confirmation with minor sacrifice of the decentralization, and can save the transaction fee [77]. Through offline communication, the transmission of clearing price, matching result and other market signals is realized within certain timing constraint [60], and the distribution system operators can also request asset transfer [37].

In [92], the off-chain payment is applied to the network optimization model of electric vehicle charging station, which not only solved the problem of privacy exposure in conventional charging system, but also addressed the drawbacks of low efficiency and high transaction cost in Bitcoin network.

The matching method in [68] mentioned above is performed using an off-blockchain solver. The solving of the matching problem is divided into two parts, the off-blockchain solver is running to find a solution on powerful computers based on the latest offers, and the smart contract on the chain only needs to verify the feasibility of the solution and to select from the multiple candidate solutions.

The idea put forward in [93] is very similar. The contract and ledger are stored separately to enhance the width of the chain from the traditional single chain to double chain, which increases reliability of the blockchain. In addition, each pair of contract block and ledger block is equipped with a high-frequency verifier, to inspect any inconsistencies and malicious manipulations. If any inconsistency is found between the recalculated ledger and the existing one, an alert will be sent to the prosumer to prevent possible data manipulations.

5.3.2 Multi-signature algorithm

Multi-signature algorithm is used to ensure transaction validity. It's basically about finding someone else to prove that neither side cheated. If the third party fails to satisfy both parties, it may be replaced. The conditions under which a transaction can be redeemed are defined using a multi-key script, the minimum M public keys can be signed to prevent theft [44,54]. If they only want to trade partial ownership, they could also adopt time locked multi-signature to refund the transaction.

Zhao [77] et al propose the concept of multi-signature wallet. The token signed by the buyer and the exchange, and is transferred from the blockchain account to the multi-signature wallet. The contract containing the order information of the buyer and the seller is sent to the exchange. The exchange classifies the orders according to the kind of energy and the delivery time, so that the orders could be settled in the same period.

5.3.3 Market transaction model

Three market transaction models are studied in [94,23].

First of all, a day ahead of trading, the market is cleared the day before it operates [95]. The products considered in the proposed day ahead market are energy and flexibility. The next day's hourly clear of demand and supply generates 24 market prices and quantities. The supply and demand participants are matched before the payment is executed. If the supply and demand do not match, a distributed flexible load is allocated to cover an inflexible supply. The load owners need to set a price and the cheapest loads will be to cover the inflexible supply. If there is still a mismatch between supply and demand after the allocation, the rest will be traded in the wholesale market. Similar ideas are applied in [47], the gap between the market and renewable energy pricing is explored firstly, then an hourly alternative energy pricing scheme is proposed that would help expand renewable energy production while reducing the carbon footprint.

A real time trading is to conduct market clearing at the same time, or close to, as operation [94-96], It generally has a certain frequency. periodically, the node sends the predicted deviation to the market and eliminates it by trading energy. In contrast to the previous settlement model, it uses a flexible payload to correct bid bias. If the flexible load in the system can successfully eliminate deviations, then the market price will be the price of the last accepted flexible load.

The final model is a system based on grid operators sending quotes and node responses. When the operator anticipates congestion, the node provides flexibility to reduce power at its own time stage, then the node sends the quantity and price it can offer, and the operator selects the tender that meets the requirements.

## 6. Category II: Consensus Mechanism

For traditional blockchain, there are several mature consensus mechanisms. When combined with energy trading, it is also likely to adopt a more traditional model or slightly modified model to meet the requirements of the system. Several common consensus mechanisms, like Proof-of-Work (PoW), Proof-of-Stake (PoS), Practical Byzantine Fault Tolerance (PBFT), Proof-of-Authority [114], Proof of Burn (PoB) and Proof of Elapsed Time (PoET) are summarized in [98,59].

*6.1. Proof-of-Work*

Since the proof-of-work mechanism was initially used in bitcoin, the most common early-stage consensus mechanism in blockchain-based energy trading was POW [60,80,81,34]. The consensus proof is achieved by solving a difficult problem, that is, finding a nonce to form a hash value satisfying the condition. Double-spending can be avoided to some extent because finding an answer requires a lot of computing resources. Other nodes can easily verify the answer. However, its weaknesses are obvious, such as the long delay to confirm a trade [94]. Additionally, POW in bitcoin cannot be directly used for energy transactions, because it can only provide integrity check for things, which is far from enough for energy transactions [68].

Later, many improved methods have been proposed on this basis. The auction-based market model is designed precisely because of the characteristics of POW mechanism in [56]. Limitations on computing resources have led miners to outsource computing to cloud computing servers. The overhead of proof-of-work is too high for mobile devices, so edge computing [90,99] and Cloud/Fog Computing [91] are used as a network support device, and two pricing models for uniform and discriminatory pricing schemes are further proposed to assist. That is, Regarding the rewards given to the miners, there are fixed rewards and variable rewards, so that the competitive factors of the miners are not only related to the calculation force, but also related to the size of the block for certification.

In [97], the data contribution frequency and energy contribution rate are specially utilized to realize consensus. The data currency and energy currency inspired by blockchain are proposed correspondingly to solve the safety problem in the interaction of electric vehicles. In [74], there are two ways as in proof-of-work mechanism. One is for local aggregators (LAGs) about data audit, it is similar to the traditional pow in Bitcoin, which generate a unique hash value of difficulty for each block. Another is for Plug-in Hybrid Electric Vehicles (PHEVs) about energy contributions, the discharge PHEV that contributes the most to the power supply is awarded with energy currency as an additional incentive. The total energy emitted is measured and recorded by smart meters, which is PHEV 's specific working proof of energy contribution.

In [100], consensus-based optimization is realized to ensure that the final demand response strategy can be trusted through the consensus among participants, so as to resist the conspiracy attack. The consensus mechanism of the combination of the consortium blockchain and the energy transaction is similar to the traditional mode in bitcoin [89]. The difference is that only a few authorized nodes are verified, and an additional leader node is added to issue the audit content to the authorized node for verification. Then the authentication results are compared. If the authentication results are consistent, the transaction will be made, and the leader node will receive additional rewards. In addition, a credit-based payment mechanism is designed to reduce the transaction confirmation time.

*6.2. Proof-of-Stake*

This is an alternative that requires little CPU power. In the proof-of-stake system, instead of competing, miners maintain a set of validators that participate in the block creation process. Each verifies that it has a stake in the grid, and the equity is used to determine the likelihood that the node adds the next block of transactions to the blockchain [25]. The system requires participants to prove the ownership of the currency. People with more money are less likely to attack networks. However, this mechanism tends to make the rich richer and the poor poorer.

*6.3. Other consensus mechanisms*

DBFT

In [82], a reputation based delegated Byzantine fault tolerance (DBFT) consensus algorithm is proposed, which reach a consensus through the cooperation of the ordinary nodes and the consensus nodes. After all non-leader consensus nodes complete this process, the appointed leader will broadcast the proposal message and send its candidate blocks to other consensus nodes. The non-leading consensus nodes compare state sets, and when different consensus nodes receive at least the same $M - f$ consensus, which means there are at least f consensus same results out of M consensus nodes, new blocks can be generated.

**7. Category III: System Optimization**

*7.1. structure optimization*

Firstly, there are many structural optimization models to improve the performance from the system design level.

In [73], a new interactive energy control model, including framework, mathematical model and pricing rules, is proposed, which can accurately match the solutions of centralized mechanisms while maintaining the privacy of the microgrid. It can be processed in parallel to reduce the time spent. In [101], an e-commerce model is developed based on the Internet of things and blockchain. Distributed autonomous companies are used as cross-departmental entities to process data and property. The system consists of three subsystems in [102], namely, the consumer system, the blockchain system and the transaction system. The users participate specifically in the consumer system and has a smart meter associated with it. Blockchain systems provide storage and are responsible for running smart contracts. The trading system provides robot advisors to advise users on the most suitable preference schemes to maximize the profits. It constitutes a private, decentralized and free energy market.

In [103], cloud computing and blockchain were still used as partners, and a new model, BCPay, is built to solve security problems in outsourcing services, achieving stability and the so-called robust equity. Analogously, in [100], a demand response model based on fog computing is proposed. The nodes in fog computing are reconstructed and the attacks can be resisted by consensus-based optimization and access control encryption. Demand response participants are divided into groups to reach consensus, and Nash equilibrium based on game theory is used to realize the random global optimization.

Considering to the impact of factors such as intermittent power generation and the grid load consumption level, it may be unable to meet the requirements from smart grid components, hence storage units are deployed in [104]. A new framework is proposed, which combines the double auction market model and non-cooperative game theory to generate a dynamic pricing mechanism that enables geographically distributed storage units to interact and trade energy. In contrast, Ilic et. al [105] focus on market energy transactions at the regional level, in which the price of energy transactions includes the necessary transmission costs but is hidden from end users. At the same time, intelligent agents are adopted to act on behalf of the end user, considering the massive interaction between the end user and the market. There are also multi-agent systems to meet different needs [67]. The role of an agent is also studied in [106], which addresses the issue of proportionately fair and two-sided energy auctions.

Moreover, for the storage problem, Wu [107] et al. designed a hybrid Blockchain architecture and a new storage mode which combines the public chain with private chain to improve the efficiency. Specifically, the public chain is used to identify data integrity, while the private chain is used to verify transaction accuracy. All blocks can be divided into management blocks and storage blocks, which are independent to each other. Users can query the data record in the management block, but they can only see the encrypted hash value instead of the original data. At the same time, the system also has the incentive mechanism driven by interests to attract more users to join the system.

*7.2. operational cost minimization*

The cost minimization involves both the individual users and the overall energy cost. For example, Lin et.al [96] mainly consider from the perspective of individual end users. The mixed integer linear programming is used to reduce energy waste and minimize energy cost. In [54], a proof-of-concept architecture based on secure private blockchain is proposed. The concept of atomic meta-transactions is introduced. The energy consumer will make a commit to pay, and after receiving the energy, an energy receipt confirmation will be generated for confirmation and connected with the smart contract. The system has proven to be able to reduce the cost, processing time, and block size of energy transactions. In [108], similarly, the operation cost of electric vehicle charging stations is prioritized, while the energy supply is not affected. It is proposed that renewable energy can be used for the operation of charging stations during the day, while the central battery pack stores the excess energy at night. In this way, the charging cost of EV owners can be minimized.

In terms of the total cost, in [49], the energy transaction between isolated island microgrids are studied, and a distributed convex optimization framework is developed for it. A cost minimization algorithm based on sub-gradient is proposed, which can expand the number of microgrids and maintain the privacy of local consumption. Hajiesmaili [109] et.al focus on solving the heterogeneity between micro grid devices. A crowd-sourcing storage system to design a centralized online algorithm is adopted, which can minimize the cost while achieving supply and demand balance. The hybrid approach combines the smart contracts of blockchain with the use of traditional computing platforms in [68].

*7.3. social welfare optimization*

The Kelly mechanism is a type of auction algorithm that allows an agent to bid separately, with bidders receiving bids allocating shares in proportion to the value of the bid. This bidding mechanism can maximize the overall benefits of all agents, namely the so-called social welfare, which is realized through effective auction [106]. This paper has also proved that double auction is easier to minimize the efficiency loss caused by price expectation than single auction, and has proposed a distributed iterative auction algorithm, which allows the selfish agent to estimate the expected price behavior through the information of previous iterations.

To achieve this, there have been many studies using different strategies. In [110], it is proposed to maximize social welfare through smart contract of blockchain and automatic initiation of fine-tuned rules. Jiao [56] et al. have designed two different cloud-based auction mechanisms for different scenes. For the constant-demand scheme, each miner bids for a fixed and limited price. While for the multi-demand scenario, the miners asked for their own needs and then issued corresponding demands freely. Both models can maximize social welfare.

In [74], according to the social welfare objective function, the optimization is calculated according to different constraints. However, detailed information is required for calculation. A reasonable mechanism is needed to extract information for calculation while ensuring user privacy. To the same end, edge computing and deep learning are used in [80] to achieve an optimal auction, which transforms based on the miner's bid and calculates payment rules, considering the authenticity and computational efficiency [81].

*7.4. Data protection in storage*

In addition to data protection in the communication process, Gai [55] et al. focused on the data stored in blocks, this part of data may be subject to link attacks or malicious data mining, hence threatens the privacy of users. In this paper, a noise-based privacy-preserving blockchain-support transaction model is proposed to generate noise to hide the transaction distribution trend. The energy sales of sellers are screened by mining the energy transaction volume. In [61], a method of grouping users is proposed. One user is randomly set to be the aggregator in each time interval. Even if the aggregators are malicious, the tampered data will be discovered by others.

*7.5. Other benefits maximization*

In addition to the cost and social welfare, there are many other performance indicators that need optimization.

In [111], an optimal charging scheduling scheme is proposed for electric vehicles by using Nondominated Sorting Genetic Algorithm (NSGA), which can maximize user satisfaction and minimize cost based on a double-objective optimization model. A novel process and an incentive mechanism are proposed to coordinate, allocate and settle energy transactions in [88], which allow producers to choose their preferred pricing strategy to maximize profits while keeping consumer costs to a minimum. There are many published studies achieving these two goals. For example, the market premium or discount based on the market price one day in advance [47], an extensible framework consisting of two independent nonlinear programming models for the coordination of residential load [64], and the pricing mechanism of the intermediate market interest rate proposed in [87].

There are also performance optimizations using game theory methods, for example, in order to solve the problem of energy management, Tushar et. al [112] for the first time take use of Shared facility controller to play an uncooperative Stackelberg game between the residential and the shared facility controller. The cost minimization and benefit maximization can be achieved. On this basis, Lee et. al [113] proposed a completely distributed energy trading mechanism, in which the seller needs to weigh the satisfaction of income and energy storage. The buyer needs to make a bid to track the seller's behavior, and the Stackelberg game is used to solve the case with multiple leaders and multiple followers, so as to realize the maximization of all micro-grid benefits. In order to maximize the utility of the operator, a method based on the contract theory is proposed to analyze the optimal contract, which can meet the individual energy consumption preference of electric vehicles and maximize the utility of the operator [82]. The two-stage Stackelberg game is used to maximize the benefits of cloud computing services and miners in a consensus process in [91].

There is also a common mode to maximize the benefits of system users by formulating appropriate auction mechanism. In [76], distributed controllers and centralized auctions are used to perform market clearing. The maximum welfare of market users can be ensured without the need for user's personal information.

## 8. Discussion and Future Directions

Increasingly more experiments show that the distributed system architecture represented by blockchain can bring greater flexibility and better performance to the energy trading market. The development of blockchain-based energy trading has made some progress and is gradually maturing. In general, the most common application form of blockchain technology in energy trading is the main framework of the system, which plays an important role in storing transaction data and ensuring security. Meanwhile, there are also those who use blockchain only as an implementation tool to achieve certain functions.

However, based on the study we presented above, there are still some issues that have not been well-addressed.

Firstly, the lack of incentive system. Many architectures only consider the basic transaction algorithm of the whole system operation, but have ignored how to stimulate the nodes to join the system, such as formulating the appropriate transaction fee model, maximizing the benefits from the perspective of end users, and so on. Certainly, it is surprising that we can see a trend that more and more systems begin to consider the interests of ordinary consumers in optimization [96,50,35]. This, on the other hand, reflects that only by focusing on the interests of individual users can the development of the power industry be stimulated to some extent.

Secondly, the improvement of consensus mechanism. At present, most systems still use POW, or the improved POW, when applying the consensus mechanism. However, many problems exist in this mechanism, such as the long time it takes to reach an agreement, and the high computational overhead. The system that we talked about earlier requires a large carbon footprint in part because of this consensus mechanism. Some of the proposed mechanisms, such as POS and PBFT, are proposed cope with the shortcomings of POW. However, they are not widely used in energy trading currently.

Thirdly, the process of transmission. There are few technologies related to route hiding. This can bring troubles to the privacy and security of users.

The last problem is the application of such systems in real life, which needs a lot of experimental and simulation tests [53]. The specific challenges include

- Lack of regulatory mechanisms. In most countries, power grid taxation is an important part of national financial income. Although transaction fees can be used to make relevant adjustments in the blockchain system, there are still needs to establish relevant systems laws and regulations. Without judicial guidance, it is hard to be widely adopted, especially by large companies.
- Implementation of the physical layer. Few systems have actually considered the implementation of hardware at the physical level, and the limitations of hardware have a great impact on the design of the system. Moreover, due to the tight network structure of the blockchain system, it is likely to cause cascading physical failures.
- Environmental challenges. Distributed ledgers require a high carbon footprint and energy consumption. Currently, few researches have been conducted on how to reduce the carbon footprint, given the high amount of data exchanged across the system. Even if the carbon footprint is not taken into account, there will still be many problems in practical applications. Additionally, the scalability of the system will also have a great impact.

In fact, apart from the application form of blockchain described in this paper, there are more extensive application forms, such as those described in [57,114], in which blockchain can be used to register and store carbon emissions, or to promote the use of renewable energy. For example, the concept of green blockchain was proposed in [115], which has been applied to another area of the energy market, namely the industrial operating systems, to manage environmental certificates, emission permits and other projects.

In the future, we should focus on these unresolved challenges. Simulation system analysis of large-scale practical applications may encounter problems. Moreover, blockchain technology has its unique advantages, and it can also be applied in other aspects of the energy market. Its application in other fields needs to be explored to generate greater benefits for the society, like the concept of green blockchain.

## 9. Conclusions

In this paper, we summarized the improvements made in each stage of the transaction in blockchain-based energy trading market. We have investigated many relevant issues, from the removal of the centralized control in distributed architecture, the solution of security and privacy issues, transaction object matching issues, the auction pricing mechanism at the time of settlement, to the cost minimization and benefit maximization of the system. Finally, we have highlighted the unsolved problems and provided insights for future directions.

**Acknowledgement**

This work is supported by Beijing Natural Science Foundation under grant 4182060.

**References**

[1]. Z Guan, J Li, L Wu, Y Zhang, J Wu, and X Du, "Achieving Efficient and Secure Data Acquisition for Cloud-supported Internet of Things in Smart Grid," IEEE Internet of Things Journal, Vol. 4, Issue 6, pp. 1934-1944, Dec. 2017.

[2]. X. Du, Y. Xiao, M. Guizani, and H. H. Chen, "An Effective Key Management Scheme for Heterogeneous Sensor Networks," Ad Hoc Networks, Elsevier, Vol. 5, Issue 1, pp 24–34, Jan. 2007.

[3]. X. Du and H. H. Chen, "Security in Wireless Sensor Networks," IEEE Wireless Communications Magazine, Vol. 15, Issue 4, pp. 60-66, Aug. 2008.

[4]. Z Guan, Y Zhang, L Wu, J Wu, Y Ma, and J Hu, "APPA: An anonymous and privacy preserving data aggregation scheme for fog-enhanced IoT," Journal of Network and Computer Applications, Vol. 125, pp. 82-92, Jan. 2019.

[5]. Z GUAN, Y ZHANG, L ZHU, L Wu, S Yu, "EFFECT: An Efficient Flexible Privacy-Preserving Data Aggregation Scheme with Authentication in Smart Grid", SCIENCE CHINA Information Sciences, Vol. 62, Issue 3, pp.1-14, Mar. 2019.

[6]. Xi Fang; Satyajayant Misra; Guoliang Xue; Dejun Yang. Smart Grid — The New and Improved Power Grid: A Survey. IEEE Communications Surveys & Tutorials. 2012, 14, 944-980.

[7]. Satoshi Nakamoto. Bitcoin: A peer-to-peer electronic cash system. Available online: https://bitcoin.org/bitcoin.pdf (accessed on 31 October 2008).

[8]. Tianyu Yang; Qinglai Guo; Xue Tai; Hongbin Sun; Boming Zhang; Wenlu Zhao; Chenhui Lin. Applying blockchain technology to decentralized operation in future energy internet. 2017 IEEE Conference on Energy Internet and Energy System Integration (EI2), Beijing, China, 26-28 Nov. 2017; DOI: 10.1109/EI2.2017.8244418.

[9]. Croman, Kyle, et al. "On scaling decentralized blockchains." International Conference on Financial Cryptography and Data Security. Springer, Berlin, Heidelberg, 2016.

[10]. Ittay Eyal; Adem Efe Gencer; Emin Gun Sirer; Robbert van Renesse. Bitcoin-NG: a scalable blockchain protocol. the Proceedings of the 13th USENIX Symposium on Networked Systems Design and Implementation (NSDI '16), Santa Clara, CA, USA, 16-18 March 2016; ISBN 978-1-931971-29-4.

[11]. David, Bernardo, et al. "Ouroboros Praos: An adaptively-secure, semi-synchronous proof-of-stake blockchain." Annual International Conference on the Theory and Applications of Cryptographic Techniques. Springer, Cham, 2018.

[12]. Zheng, Z., Xie, S., Dai, H. N., Chen, X., & Wang, H. (2018). Blockchain challenges and opportunities: A survey. International Journal of Web and Grid Services, 14(4), 352-375.

[13]. Guy Zyskind; Oz Nathan; Alex 'Sandy' Pentland. Decentralizing Privacy: Using Blockchain to Protect Personal Data. 2015 IEEE SPW, San Jose, CA, USA, 21-22 May 2015; DOI: 10.1109/SPW.2015.27.

[14]. Ahmed Kosba; Andrew Miller; Elaine Shi; Zikai Wen; Charalampos Papamanthou. Hawk: The Blockchain Model of Cryptography and Privacy-Preserving Smart Contracts. IEEE Symposium on Security and Privacy, San Jose, CA, USA, 22-26 May 2016; DOI 10.1109/SP.2016.55.

[15]. Konstantinos Christidis; Michael Devetsikiotis. Blockchains and Smart Contracts for the Internet of Things. IEEE Access. 2016, 4, 2292-2303.

[16]. Lijing Zhou; Licheng Wang; Yiru Sun; Pin Lv. BeeKeeper: A Blockchain-based IoT System with Secure Storage and Homomorphic Computation. IEEE Access. 2018, 6, 43472-43488.

[17]. Z Li, S Mohammad, X Liu, "Cyber-secure decentralized energy management for IoT-enabled active distribution networks", J. Mod. Power Syst. Clean Energy, Vol. 6, pp. 900-917, 2018.

[18]. S Hu, C Cai, Q Wang, C Wang, X Luo, K Ren, "Searching an Encrypted Cloud Meets Blockchain: A Decentralized, Reliable and Fair Realization", IEEE Conference on Computer Communications (Infocom 2018), Honolulu, HI, USA, pp. 16-19, April 2018.

[19]. L Huang, G Zhang, S Yu, A Fu, and J Yearwood, "Customized Data Sharing Scheme based on Blockchain and Weighted Attribute", Globecom 2018, pp. 1-6, Abu Dhabi, UAE, 2018.

[20]. C Park, and Y Taeseok, "Comparative review and discussion on P2P electricity trading." Energy Procedia, Vol. 128, pp. 3-9, 2017.

[21]. Juhar Abdella, Khaled Shuaib, "Peer to Peer Disributed Energy Trading in Smart Grids A Survey", Energies. 2018, 11, 1-22.

[22]. Tiago Sousa; Tiago Soares; Pierre Pinson; Fabio Moret; Thomas Baroche; Etienne Sorin. Peer-to-pee and community-based markets A comprehensive review. Computers and Society. 2018, arXiv:1810.09859.

[23]. Mohsen Khorasany; Yateendra Mishra; Gerard Ledwich. Market framework for local energy trading a review of potential designs and market clearing approaches. IET Generation, Transmission & Distribution. 2018, 12, 5899 – 5908.

[24]. J Murkin, R Chitchyan, A Byrne, "Enabling peer-to-peer electricity trading", 4th International Conference on ICT for Sustainability (ICT4S 2016), pp. 234-235, 2016.

[25]. C Pop, T Cioara, M Antal, I Anghel, I Salomie, M Bertoncini, "Blockchain Based Decentralized Management of Demand Response Programs in Smart Energy Grids", Sensors. 2018, 18, 162-162.

[26]. J J.Sikorski, J Haughton, M Kraft, "Blockchain technology in the chemical industry Machine-to-machine electricity market", Applied Energy. 2017, 195, 234-246.

[27]. Khaled Shuaib; Juhar Ahmed Abdella; Farag Sallabi; Mohammed Abdel-Hafez. Using Blockchains to Secure Distributed Energy Exchange. 2018 5th International Conference on Control, Decision and Information Technologies (CoDIT), Thessaloniki, Greece, 10-13 April 2018, DOI: 10.1109/CoDIT.2018.8394815.

[28]. Javier Matamoros; David Gregoratti; Mischa Dohler. Microgrids Energy Trading in Islanding Mode. 2012 IEEE Third International Conference on Smart Grid Communications (SmartGridComm), Tainan, Taiwan, 5-8 Nov. 2012; DOI: 10.1109/SmartGridComm.2012.6485958.

[29]. Xingzheng Zhu; Ka-Cheong Leung. An adaptive distributed power scheduling algorithm for renewable Microgrid cooperation. 2017 IEEE International Conference on Communications (ICC), Paris, France, 21-25 May 2017; DOI: 10.1109/ICC.2017.7996940.

[30]. Yu Wang; Shiwen Mao; R. M. Nelms. On Hierarchical Power Scheduling for the Macrogrid and Cooperative Microgrids. IEEE TII. 2015, 11, 1574-1584.

[31]. Esther Mengelkamp; Benedikt Notheisen; Carolin Beer; David Dauer; Christof Weinhardt. A blockchain-based smart grid towards sustainable local energy markets. Computer Science - Research and Development. 2018 , 33, 207-214.

[32]. Ioannis Kounelis; Gary Steri; Raimondo Giuliani; Dimitrios Geneiatakis; Ricardo Neisse; Igor Nai-Fovino. Fostering consumers' energy market through smart contracts. 2017 International Conference in Energy and

Sustainability in Small Developing Economies (ES2DE), Funchal, Portugal, 10-12 July 2017; DOI: 10.1109/ES2DE.2017.8015343.

[33]. Chaoyue Gao; Yanchao Ji; Jianze Wang; Xiangyv Sai. Application of Blockchain Technology in Peer-to-Peer Transaction of Photovoltaic Power Generation. 2018 2nd IEEE Advanced Information Management,Communicates,Electronic and Automation Control Conference (IMCEC), Xi'an, China, 25-27 May 2018; DOI: 10.1109/IMCEC.2018.8469363.

[34]. Sandi Rahmadika; Diena Rauda Ramdania; Maisevli Harika. Security Analysis on the Decentralized Energy Trading System Using Blockchain Technology. JOIN. 2018, 3, 44-47.

[35]. Junyeon Hwang; Myeong-in Choi; Tacklim Lee; Seonki Jeon; Seunghwan Kim; Sounghoan Park; Sehyun Park. Energy Prosumer Business Model Using Blockchain System to Ensure Transparency and Safety. Energy Procedia. 2017, 141, 194-198.

[36]. Bent Richter; Esther Mengelkamp; Christof Weinhardt. Maturity of Blockchain Technology in Local Electricity Markets. 2018 15th International Conference on the European Energy Market (EEM), Lodz, Poland, 27-29 June 2018; DOI: 10.1109/EEM.2018.8469955.

[37]. X. Du, Y. Xiao, S. Ci, M. Guizani, and H. H. Chen, "A Routing-Driven Key Management Scheme for Heterogeneous Sensor Networks," in Proc. of IEEE International Conference on Communications (ICC 2007), Glasgow, Scotland, June 2007.

[38]. Ian Miers; Christina Garman; Matthew Green; Aviel D. Rubin. Zerocoin: Anonymous Distributed E-Cash from Bitcoin. 2013 IEEE Symposium on Security and Privacy, Berkeley, CA, USA, 19-22 May 2013; DOI: 10.1109/SP.2013.34.

[39]. Y. Xiao, V. Rayi, B. Sun, X. Du, F. Hu, and M. Galloway, "A Survey of Key Management Schemes in Wireless Sensor Networks," Journal of Computer Communications, Vol. 30, Issue 11-12, pp. 2314-2341, Sept. 2007.

[40]. Y. Xiao, X. Du, J. Zhang, and S. Guizani, "Internet Protocol Television (IPTV): the Killer Application for the Next Generation Internet," IEEE Communications Magazine, Vol. 45, No. 11, pp. 126–134, Nov. 2007.

[41]. X. Du, M. Guizani, Y. Xiao and H. H. Chen, Transactions papers, "A Routing-Driven Elliptic Curve Cryptography based Key Management Scheme for Heterogeneous Sensor Networks," IEEE Transactions on Wireless Communications, Vol. 8, No. 3, pp. 1223 - 1229, March 2009.

[42]. J. Wu, M. Dong, K. Ota, J. Li, Z. Guan, "Big Data Analysis-Based Secure Cluster Management for Optimized Control Plane in Software-Defined Networks," IEEE Transactions on Network and Service Management, vol. 15, no. 1, pp. 27-38, 2018.

[43]. J. Wu, M. Dong, K. Ota, J. Li, Z. Guan, "FCSS: Fog-Computing-based Content-Aware Filtering for Security Services in Information-Centric Social Networks," IEEE Transactions on Emerging Topics in Computing, Vol. PP, Issue 99, pp. 1-1, 2018.

[44]. Nurzhan Zhumabekuly Aitzhan; Davor Svetinovic. Security and Privacy in Decentralized Energy Trading Through Multi-Signatures, Blockchain and Anonymous Messaging Streams. IEEE TDSC. 2018, 15, 840-852.

[45]. Jiani Wu; Nguyen Khoi Tran. Application of Blockchain Technology in Sustainable Energy Systems An Overview. Sustainability. 2018, 10, DOI:10.3390/su10093067.

[46]. Jian Wang; Qianggang Wang; Niancheng Zhou; Yuan Chi. A Novel Electricity Transaction Mode of Microgrids Based on Blockchain and Continuous Double Auction. Energies. 2017, 10, DOI:10.3390/en10121971.

[47]. Adrian Ettlin. Dynamic Modeling of Peer-to-Peer Power Market Making. 2018 15th International Conference on the European Energy Market (EEM), Lodz, Poland, 27-29 June 2018; DOI: 10.1109/EEM.2018.8469775.

[48]. Michael Mylrea; Sri Nikhil Gupta Gourisetti. Blockchain for smart grid resilience Exchanging distributed energy at speed, scale and security. 2017 Resilience Week (RWS), Wilmington, DE, USA, 18-22 Sept. 2017; DOI: 10.1109/RWEEK.2017.8088642.

[49]. David Gregoratti; Javier Matamoros. Distributed Energy Trading: The Multiple-Microgrid Case. IEEE TIE. 2015, 62, 2551-2559.

[50]. Olamide Jogunola; Augustine Ikpehai; Kelvin Anoh; Bamidele Adebisi; Mohammad Hammoudeh; Haris Gacanin; Georgina Harris. Comparative Analysis of P2P Architectures for Energy Trading and Sharing. Energies. 2017, 11, 1-20.

[51]. Mohamed Amine Ferrag; Makhlouf Derdour; Mithun Mukherjee; Abdelouahid Derhab; Leandros Maglaras; Helge Janicke. Blockchain Technologies for the Internet of Things Research Issues and Challenges. IEEE JIOT. 2018, DOI: 10.1109/JIOT.2018.2882794.
[52]. Thomas Morstyn; Niall Farrell; Sarah J. Darby; Malcolm D. McCulloch. Using peer-to-peer energy-trading platforms to incentivize prosumers to form federated power plants. Nature Energy. 2018, 3, 94-101.
[53]. Van Hoa Nguyen; Yvon Besanger; Quoc Tuan Tran; Minh Tri Le. On the Applicability of Distributed Ledger Architectures to Peer-to-Peer Energy Trading Framework. 2018 IEEE International Conference on Environment and Electrical Engineering and 2018 IEEE Industrial and Commercial Power Systems Europe (EEEIC / I&CPS Europe), Palermo, Italy, 12-15 June 2018; DOI: 10.1109/EEEIC.2018.8494446.
[54]. Dorri Ali; Hill Ambrose; Kanhere Salil S; Jurdak Raja; Luo Fengji; Dong Zhao Yang. Peer-to-Peer EnergyTrade A Distributed Private Energy Trading Platform. Cryptography and Security. 2018, arXiv:1812.08315.
[55]. K Gai, Yulu Wu, L Zhu, M Qiu, M Shen, "Privacy-preserving Energy Trading Using Consortium Blockchain in Smart Grid", IEEE Transactions on Industrial Informatics, Vol. PP, no. 99, pp. 1-1, Jan. 2019.
[56]. Yutao Jiao; Ping Wang; Dusit Niyato; Kongrath Suankaewmanee. Auction Mechanisms in Cloud/Fog Computing Resource Allocation for Public Blockchain Networks. Computer Science and Game Theory. 2018, arXiv:1804.09961.
[57]. VladaBrilliantova; Thomas WolfgangThurner. Blockchain and the future of energy. Technology in Society. 2018, DOI:10.1016/j.techsoc.2018.11.001.
[58]. Perukrishnen Vytelingum; Sarvapali D. Ramchurn; Thomas D. Voice; Alex Rogers; Nicholas R. Jennings. Trading agents for the smart electricity grid. AAMAS '10 Proceedings of the 9th International Conference on Autonomous Agents and Multiagent Systems. 2010, 1, 897-904.
[59]. Aron Laszka; Abhishek Dubey; Michael Walker; Doug Schmidt. Providing Privacy, Safety, and Security in IoT-Based Transactive Energy Systems using Distributed Ledgers. IoT '17, Linz, Austria, 22–25 October 2017; DOI:10.1145/3131542.3131562.
[60]. Karla Kvaternik; Aron Laszka; Michael Walker; Douglas Schmidt; Monika Sturm; Martin lehofer, Abhishek Dubey. Privacy-Preserving Platform for Transactive Energy Systems. 2018, arXiv:1709.09597.
[61]. Z Guan, G Si, X Zhang, L Wu, N Guizani, X Du, and Y Ma. Privacy-preserving and Efficient Aggregation based on Blockchain for Power Grid Communications in Smart Communities. IEEE Communications Magazine, Jul. 2018; Vol. 56, Issue 7, pp. 82-88.
[62]. Yanan Sun; Lutz Lampe; Vincent W. S. Wong. Smart Meter Privacy: Exploiting the Potential of Household Energy Storage Units. IEEE JIOT. 2017, 5, 69-78.
[63]. Aysajan Abidin; Abdelrahaman Aly; Sara Cleemput; Mustafa A. Mustafa. Secure and Privacy-Friendly Local Electricity Trading and Billing in Smart Grid. 2018, arXiv:1801.08354.
[64]. Armin Ghasem Azar; Hamidreza Nazaripouya; Behnam Khaki; Chi-Cheng Chu; Rajit Gadh; Rune Hylsberg Jacobsen. A Non-Cooperative Framework for Coordinating a Neighborhood of Distributed Prosumers. IEEE TII. 2018, DOI: 10.1109/TII.2018.2867748.
[65]. Yu Yong; Ding Yujie; Zhao Yanqi; Li Yannan; Du Xiaojiang; Guizani Mohsen. LRCoin, Leakage-resilient Cryptocurrency Based on Bitcoin for Data Trading in IoT. IEEE JOIT. 2018, DOI: 10.1109/JIOT.2018.2878406.
[66]. F Yucel, E Bulut, K Akkaya. "Privacy Preserving Distributed Stable Matching of Electric Vehicles and Charge Suppliers", IEEE VTC.
[67]. H. S. V. S. Kumar Nunna; Suryanarayana Doolla. Multiagent-Based Distributed-Energy-Resource Management for Intelligent Microgrids. IEEE TIE. 2013, 60, 1678-1687.
[68]. Aron Laszka, Abhishek Dubey, Scott Eisele, Michael Walker, Karla Kvaternik. Design and Implementation of Safe and Private Forward-Trading Platform for IoT-Based Transactive Microgrids. Available online: http://www.isis.vanderbilt.edu/node/4868. (accessed on 18 December 2017)
[69]. Weifeng Zhong; Chao Yang; Kan Xie; Shengli Xie; Yan Zhang. ADMM-Based Distributed Auction Mechanism for Energy Hub Scheduling in Smart Buildings. IEEE Access. 2018, 6, 45635-45645.
[70]. Jangkyum Kim; Joohyung Lee; Sangdon Park; Jun Kyun Choi. Battery Wear Model based Energy Trading in Electric Vehicles A Naive Auction Model and a Market Analysis. IEEE TII. 2018, DOI: 10.1109/TII.2018.2883655.
[71]. Bodhisattwa P. Majumder; M. Nazif Faqiry; Sanjoy Das; Anil Pahwa. An efficient iterative double auction for energy trading in microgrids. 2014 IEEE Symposium on Computational Intelligence Applications in Smart Grid (CIASG), Orlando, FL, USA, 9-12 Dec. 2014; DOI: 10.1109/CIASG.2014.7011556.

[72]. Bhuvaneswari Ramachandran; Sanjeev K. Srivastava; Chris S. Edrington; David A. Cartes. An Inelligent Auction Scheme for Smart Grid Market Using a Hybrid Immune Algorithm. IEEE TIE. 2011, 58, 4603-4612.

[73]. W Liu, J Zhan, C.Y. Chung, "A Novel Transactive Energy Control Mechanism for Collaborative Networked Microgrids", IEEE TPWRS. 2018, DOI: 10.1109/TPWRS.2018.2881251.

[74]. Jiawen Kang, Rong Yu, Xumin Huang, Sabita Maharjan, Yan Zhang, Ekram Hossain, "Enabling Localized Peer-to-Peer Electricity Trading Among Plug-in Hybrid Electric Vehicles Using Consortium Blockchains", IEEE TII. 2017, 13, 3154-3164.

[75]. Jaysson Guerrero; Archie C. Chapman; Gregor Verbič. Decentralized P2P Energy Trading under Network Constraints in a Low-Voltage Network. IEEE TSG. 2018, DOI: 10.1109/TSG.2018.2878445.

[76]. Mohsen Khorasany; Yateendra Mishra; Gerard Ledwich. Auction based energy trading in transactive energy market with active participation of prosumers and consumers. 2017 Australasian Universities Power Engineering Conference (AUPEC), Melbourne, VIC, Australia, 19-22 Nov. 2017; DOI: 10.1109/AUPEC.2017.8282470.

[77]. Shengnan Zhao; Beibei Wang; Yachao Li; Yang Li. Integrated Energy Transaction Mechanisms Based on Blockchain Technology. Energies. 2018, 11, 1-19.

[78]. Wayes Tushar; Chau Yuen; Hamed Mohsenian-Rad; Tapan Saha; H. Vincent Poor; Kristin L Wood. Transforming Energy Networks via Peer to Peer Energy Trading: Potential of Game Theoretic Approaches. IEEE Signal Processing Magazine. 2018, 35, 90-111.

[79]. Chenghua Zhang; Jianzhong Wu; Yue Zhou; Meng Cheng; Chao Long. Peer-to-Peer energy trading in a Microgrid. Applied Energy. 2018, 220, 1-12.

[80]. Nguyen Cong Luong; Zehui Xiong; Ping Wang; Dusit Niyato. Optimal Auction For Edge Computing Resource Management. 2018 IEEE International Conference on Communications (ICC), Kansas City, MO, USA, 20-24 May 2018, DOI: 10.1109/ICC.2018.8422743.

[81]. Yutao Jiao; Ping Wang; Dusit Niyato; Zehui Xiong. Social Welfare Maximization Auction in Edge Computing Resource Allocation for Mobile Blockchain. 2018 IEEE International Conference on Communications (ICC), Kansas City, MO, USA, 20-24 May 2018, DOI: 10.1109/ICC.2018.8422632.

[82]. Zhou Su, Yuntao Wang, Qichao Xu, Minrui Fei, Yu-Chu Tian, Ning Zhang, "A secure charging scheme for electric vehicles with smart communities in energy blockchain", IEEE Internet of Things Journal, Vol. PP, no. 99, pp. 1-1, Sep. 2018.

[83]. Ning Wang; Weisheng Xu; Zhiyu Xu; Weihui Shao. Peer-to-Peer Energy Trading among Microgrids with Multidimensional Willingness. Energies. 2018, 11, 1-12.

[84]. Cong Nam Truong; Michael Schimpe; Uli Bürger; Holger C. Hesse; Andreas Jossen. Multi-Use of Stationary Battery Storage Systems with Blockchain Based Markets. Energy Procedia. 2018, 155, 3-16.

[85]. Woongsup Lee; Lin Xiang; Robert Schober; Vincent W. S. Wong. Direct Electricity Trading in Smart Grid A Coalitional Game Analysis. IEEE JSAC. 2014, 32, 1398-1411.

[86]. Chao Long; Jianzhong Wu; Yue Zhou; Nick Jenkins. Peer-to-peer energy sharing through a two-stage aggregated battery control in a community Microgrid. Applied Energy. 2018, 226, 261-276.

[87]. Wayes Tushar; Tapan Kumar Saha; Chau Yuen; Paul Liddell; Richard Bean; H. Vincent Poor. Peer-to-Peer Energy Trading With Sustainable User Participation A Game Theoretic Approach. IEEE Access. 2018, 6, 62932-62943.

[88]. Arne Meeuw. Design, Implementation, and Evaluation of a Blockchain-enabled Multi-Energy Transaction System for District Energy Systems. Master's Thesis. Swiss Federal Institute of Technology (ETH), Zurich, April 2018.

[89]. Zhetao Li; Jiawen Kang; Rong Yu; Dongdong Ye; Qingyong Deng; Yan Zhang. Consortium Blockchain for Secure Energy Trading in Industrial Internet of Things. IEEE TII. 2018, 14, 3690-3700.

[90]. Xiong, Z., Feng, S., Niyato, D., Wang, P., & Han, Z. Edge computing resource management and pricing for mobile blockchain. arXiv preprint arXiv:1710.01567, 2017.

[91]. Zehui Xiong; Shaohan Feng; Wenbo Wang; Dusit Niyato; Ping Wang; Zhu Han. Cloud/Fog Computing Resource Management and Pricing for Blockchain Networks. IEEE JIOT. 2018, DOI: 10.1109/JIOT.2018.2871706.

[92]. Enes Erdin, Mumin Cebe, Kemal Akkaya, Senay Solak, Eyuphan Bulut, Selcuk Uluagac, "Building a Private Bitcoin-based Payment Network among Electric Vehicles and Charging Stations", Blockchain-2018.

[93]. Fengji Luo; Zhao Yang Dong; Gaoqi Liang; Junichi Murata; Zhao Xu. A Distributed Electricity Trading System in Active Distribution Networks Based on Multi-Agent Coalition and Blockchain. IEEE TPERS. 2018, DOI 10.1109/TPWRS.2018.2876612.

[94]. F Blom, F Hossein, "On the Scalability of Blockchain-Supported Local Energy Markets." 2018 International Conference on Smart Energy Systems and Technologies (SEST). IEEE, Sevilla, Spain, pp. 1-6, 2018.

[95]. Fredrik Blom; Hossein Farahmand. On the Scalability of Blockchain-Supported Local Energy Markets. 2018 International Conference on Smart Energy Systems and Technologies (SEST), Sevilla, Spain, 10-12 Sept. 2018; DOI: 10.1109/SEST.2018.8495882.

[96]. Chun-Cheng Lin; Der-Jiunn Deng; Chih-Chi Kuo; Yu-Lin Liang. Optimal Charging Control of Energy Storage and Electric Vehicle of an Individual in the Internet of Energy With Energy Trading. IEEE TII. 2018, 14, 2570-2578.

[97]. Hong Liu; Yan Zhang; Tao Yang. Blockchain-Enabled Security in Electric Vehicles Cloud and Edge Computing. IEEE Network. 2018, 32, 78-83.

[98]. Maria Luisa Di Silvestre; Pierluigi Gallo; Mariano Giuseppe Ippolito; Eleonora Riva Sanseverino; Gaetano Zizzo. A Technical Approach to the Energy Blockchain in Microgrids. IEEE TII. 2018, 14, 4792 – 4803.

[99]. Yuan Wu; Xiangxu Chen; Jiajun Shi; Kejie Ni; Liping Qian; Liang Huang; Kuan Zhang. Optimal Computational Power Allocation in Multi-Access Mobile Edge Computing for Blockchain. Sensors. 2018, 18, DOI.org/10.3390/s18103472.

[100]. Gaolei Li; Jun Wu; Jianhua Li; Zhitao Guan; Longhua Guo. Fog Computing-Enabled Secure Demand Response for Internet of Energy Against Collusion Attacks Using Consensus and ACE. IEEE Access. 2018, 6,11278-11288.

[101]. Yu Zhang; Jiangtao Wen. The IoT electric business model: Using blockchain technology for the internet of things. 2016, Peer-to-Peer Networking and Applications. 2017, 10, 983-994. 102.

[102]. Katiuscia Mannaro; Andrea Pinna; Michele Marchesi. Crypto-trading: Blockchain-oriented energy market. 2017 AEIT International Annual Conference, Cagliari, Italy, 20-22 Sept. 2017; DOI: 10.23919/AEIT.2017.8240547.

[103]. Yinghui Zhang; Robert H.Deng; Ximeng Liu; Dong Zheng. Blockchain based efficient and robust fair payment for outsourcing services in cloud computing. Information Sciences. 2018, 462, 262-277.

[104]. Yunpeng Wang; Walid Saad; Zhu Han; H. Vincent Poor; Tamer Başar.A Game-Theoretic Approach to Energy Trading in the Smart Grid. IEEE TSG. 2014, 5, 1439-1450.

[105]. Dejan Ilic; Per Goncalves Da Silva; Stamatis Karnouskos; Martin Griesemer. An energy market for trading electricity in smart grid neighbourhoods. 2012 6th IEEE International Conference on Digital Ecosystems and Technologies (DEST), Campione d'Italia, Italy, 18-20 June 2012; DOI: 10.1109/DEST.2012.6227918.

[106]. M. Nazif Faqiry; Sanjoy Das. Double-Sided Energy Auction in Microgrid: Equilibrium Under Price Anticipation. IEEE Access. 2016, 4, 3794-3805.

[107]. Lijun Wu; Kun Meng; Shuo Xu; Shuqin Li; Meng Ding; Yanfeng Suo. Democratic Centralism A Hybrid Blockchain Architecture and Its Applications in Energy Internet. 2017 IEEE International Conference on Energy Internet (ICEI), Beijing, China, 17-21 April 2017; DOI: 10.1109/ICEI.2017.38.

[108]. Lindiwe Bokopane; Kanzumba Kusakana; Herman Jacobus Vermaak. Energy Management of a Grid-Intergrated Hybrid Peer-to-Peer Renewable Charging Station for Electric Vehicles. 2018 Open Innovations Conference (OI), Johannesburg, South Africa, 3-5 Oct. 2018; DOI: 10.1109/OI.2018.8535881.

[109]. Mohammad H. Hajiesmaili; Minghua Chen; Enrique Mallada; Chi-Kin Chau. Crowd-Sourced Storage-Assisted Demand Response in Microgrids. e-Energy '17, Shatin, Hong Kong, 16-19 May 2017; DOI:10.1145/3077839.3077841.

[110]. Ilak Perica; Rajšl Ivan; Zmijarević Zlatko; Herenčić Lin; Krajcar Slavko. Decentralised electricity trading in the microgrid Implementation of decentralized Peer-to-Peer Concept for Electricity Trading (P2PCET). 11th Mediterranean Conference on Power Generation, Transmission, Distribution and Energy Conversion, Dubrovnik, Hrvatska, 12-15 November 2018; https://www.researchgate.net/publication/329093264.

[111]. Xiaohong Huang; Yong Zhang; Dandan Li; Lu Han. An optimal scheduling algorithm for hybrid EV charging scenario using consortium blockchains. Future Generation Computer Systems. 2019, 91, 555-562.

[112]. Wayes Tushar; Bo Chai; Chau Yuen; David B. Smith; Kristin L. Wood; Zaiyue Yang; H. Vincent Poor. Three-Party Energy Management With Distributed Energy Resources in Smart Grid. IEEE TIE. 2015, 62, 2487-2498.

[113]. Joohyung Lee; Jun Guo; Jun Kyun Choi; Moshe Zukerman. Distributed Energy Trading in Microgrids A Game-Theoretic Model and Its Equilibrium Analysis. IEEE TIE. 2015, 62, 3524-3533.

[114]. A Goranović, M Meisel, L Fotiadis, S Wilker, A Treytl, T Sauter, "Blockchain applications in microgrids an overview of current projects and concepts." Industrial Electronics Society, IECON 2017-43rd Annual Conference of the IEEE. IEEE, pp. 6153-6158, Beijing China, 2017.

[115]. F. Imbault; M. Swiatek; R. de Beaufort; R. Plana. The green blockchain Managing decentralized energy production and consumption. 2017 IEEE International Conference on Environment and Electrical Engineering and 2017 IEEE Industrial and Commercial Power Systems Europe (EEEIC / I&CPS Europe), Milan, Italy, 6-9 June 2017; DOI: 10.1109/EEEIC.2017.7977613.